\documentstyle[twoside,fleqn,espcrc2,epsf]{article}

\title{Heavy quark potential in the instanton liquid model}

\author{R.~C.~Brower\address{Department of Physics, Boston University,
		Boston, MA 02215},
	D.~Chen\address{Center for Theoretical Physics,
		Massachusetts Institute of Technology,
		Cambridge, MA 02139}\thanks{Presented by D.~Chen.  Work
		supported by the Department of Energy under cooperative
		research agreement DE--FC02--94 ER40818.},
	J.~W.~Negele$^{\rm b}$ and
	E.~Shuryak\address{Department of Physics and Astronomy,
		State University of New York,
		Stony Brook, NY 11794}
}
       
\begin{document}

\begin{abstract}
We study the heavy quark potential in the instanton liquid model by
carefully measuring Wilson loops out to a distance of order 3$fm$. A
random instanton ensemble with a fixed radius $\rho$ = 1/3$fm$ generates
a potential $V(R)$ growing very slowly at large $R$.  In contrast,
a more realistic size distribution growing as $\rho^6$ at small $\rho$
and decaying as $\rho^{-5}$ at large $\rho$, leads to a potential
which grows linearly with $R$.  The string tension, however, is only
about 1/10 of the phenomenological value.
\end{abstract}

\maketitle

\section{INTRODUCTION}
There is growing evidence that instantons play a very important role
in QCD.  The development of the Interacting Instanton Liquid Model
\cite{Shuryak98} allows one to calculate nonperturbatively to all
orders in the `t Hooft interaction, and results show that it correctly
generates the quark condensate and most salient properties of light
hadrons.  As reviewed at this conference \cite{Negele98}, the essential
features and parameters of this model have now been confirmed on the lattice,
and there is strong lattice evidence that instantons and their associated
zero modes play a significant role in hadron structure.

Motivated by a provocative result by Fuku-shima {\it et} 
{\it al.}~\cite{Fukushima98} reporting a string tension
in the instanton liquid
model close to the phenomenological value, we have undertaken a careful
calculation of the instanton induced heavy quark potential.

\section{SIMULATION DETAILS}
The gauge field of a single instanton in $SU(2)$ in the singular gauge
(centered at the origin) is given by
\begin{equation}
A_{\mu}(x) = 2 \tau^{a} \bar{\eta}_{a\mu\nu}
	\frac{x_\nu}{x^2} \frac{\rho^2}{x^2+\rho^2},
\end{equation}
where $\bar{\eta}_{a\mu\nu}$ is the `t Hooft symbol.  Replacement of
$\bar{\eta}_{a\mu\nu}$ by $\eta_{a\mu\nu}$ yields the solution for
an anti-instanton.  The multi-instanton gauge configuration in the dilute
instanton liquid model is generated by a linear superposition of
the fields from a distribution of instantons and anti-instantons.
Each instanton and anti-instanton has a random center location and 
is embedded in $SU(3)$ with a random color orientation.

We consider two distributions of the instanton size, $\rho$, for $SU(3)$:
(1) a fixed size distribution, with $\rho = 1/3 fm$;
and (2) a phenomenological distribution, with $D(\rho) \sim \rho^6$
at small $\rho$ as expected from the dilute instanton gas approximation
and $D(\rho) \sim \rho^{-5}$ at large $\rho$ as would occur if the
perturbative running of the coupling constant $g$ were frozen
at sufficiently large $\rho$.  We use the parameterization by
Shuryak\cite{Shuryak98} for this variable size distribution,
\begin{equation}
D(\rho) = C \frac{\rho^6}{(\rho_0^{3.5} + \rho^{3.5})^\frac{11}{3.5}}.
	\label{eq:size}
\end{equation}
We choose $\rho_0$ so that the average instanton size $\bar{\rho} = 1/3 fm$
and $C$ is a normalization constant.

\begin{figure}[t]
\vspace{-0.2in}
\epsfxsize=\hsize
%\epsfbox[25 18 587 524]{fix-n1_0-eff.ps}
\epsffile{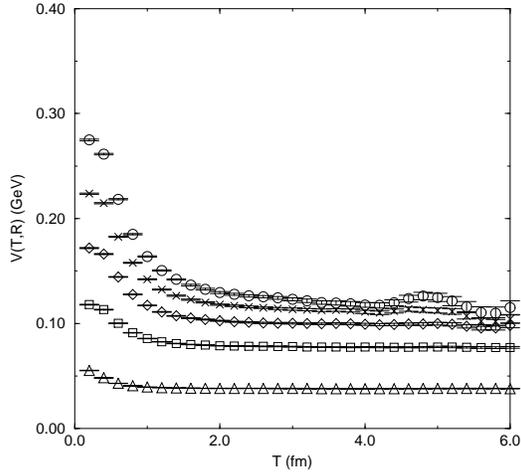}
\vspace{-0.4in}
\caption{$V(T,R)$ as a function of $T$ for the fixed instanton size 
distribution $\rho = 1/3fm$ at an instanton number density $n = 1.0fm^{-4}$.
Each data set from bottom to top corresponds to $R=$ $0.4fm$, $0.8fm$, $1.2fm$,
$1.6fm$ and $2.0fm$, respectively.}
\label{fig:fixeff}
\end{figure}

We generate ensembles of multi-instanton configurations 
in an open box of size
$(L_x, L_y, L_z, L_t)$ $=$ $(13.6fm, 7.2fm, 7.2fm, 20fm)$.
Wilson loops
\begin{equation}
W(R,T) = \rm{tr} P \exp[i \oint_{R \times T} A_{\mu} dx_{\mu}]
\end{equation}
are measured in a $6.4 fm \times 12.8 fm$ rectangle in the central
$yz$ plane.  This leaves a distance of at least $3.6 fm$ or more
than 10 times the average instanton size $\bar{\rho}$ from the 
measured Wilson loop to the 4-$d$ boundary so that the edge effects
from the open boundary are small.  The largest Wilson loop is
$3.2fm \times 6.4fm$ in size.  The discretization $\Delta x$ in the
path ordered integral is $0.05 fm$, which is about $1/6$ of 
the average instanton size.

We study three different instanton number densities, $n = N/V = 0.5 fm^{-4}$,
$1.0 fm^{-4}$ and $1.5 fm^{-4}$ for both fixed and variable instanton size
distributions.  The number of gauge configurations studied for each
different $n$ is 1600, 8000 and 1600, respectively.

\begin{figure}[t]
\vspace{-0.2in}
\epsfxsize=\hsize
%\epsfbox[12 12 576 524]{var-n1_0-eff.ps}
\epsffile{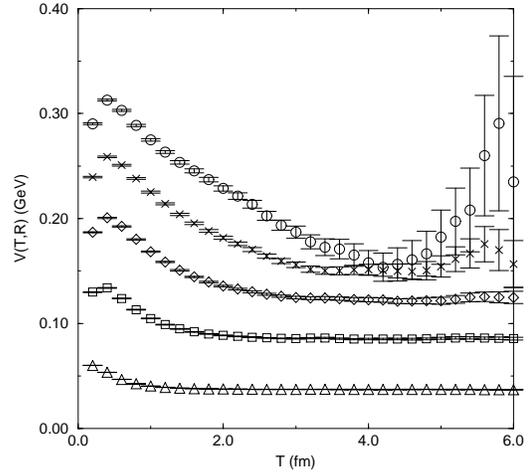}
\vspace{-0.4in}
\caption{$V(T,R)$ as a function of $T$ for a variable instanton size 
distribution described in Eq.~(\ref{eq:size}) at an instanton number
density $n = 1.0fm^{-4}$.
Each data set from bottom to top corresponds to $R=$ $0.4fm$, $0.8fm$, $1.2fm$,
$1.6fm$ and $2.0fm$, respectively.}
\label{fig:vareff}
\end{figure}

\section{RESULTS}
The fundamental difficulty in this calculation is measuring with adequate
statistical accuracy the large Wilson loops required to determine the
potential at large distances.  We determine the heavy quark potential $V(R)$
from rectangular loops of different time extent as follows:
\begin{eqnarray}
V(T,R) & = & - \frac{1}{\Delta} \log \frac{W(R, T+\Delta)}{W(R,T)}
  \label{eq:V},\\
V(R) & = & \lim_{T \rightarrow \infty} V(T,R).
\end{eqnarray}
To obtain the most accurate measurement of the potential, we plot $V(T,R)$
vs.~T and find the plateau corresponding to $V(R)$.

Figures \ref{fig:fixeff} and \ref{fig:vareff} show such plots at different $R$
for both fixed and variable instanton size distributions at an instanton number
density $n = 1.0fm^{-4}$.  For the fixed instanton size distribution, 
$\rho = 1/3fm$, plateaus for $V(T,R)$ set in at about $T > (1\sim2) R$.
We have good plateaus for all $R \leq 3.2 fm$ at $n=1.0fm^{-4}$.  In contrast,
for the variable instanton size distribution, not only do the plateaus set in
at a larger $T > (2-3)R$, but also the statistical fluctuations are much
larger at large $R$, especially for $R \geq 2.0fm$.  It is clear that this
behavior of the variable size distribution arises from large
instantons in the tail of the distribution.  Even with an ensemble as large
as 8000 configurations, we are only able to
obtain a potential $V(R)$ for $R \leq 2.0fm$ at $n = 1.0 fm^{-4}$.

\begin{figure}[t]
\vspace{-0.2in}
\epsfxsize=\hsize
%\epsfbox[25 18 587 524]{pot-fix.ps}
\epsffile{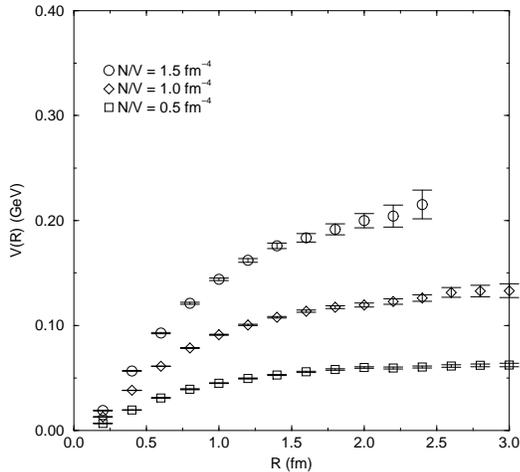}
\vspace{-0.4in}
\caption{Heavy quark potential in the instanton liquid model for a fixed
instanton size $\rho=1/3fm$.  Three different instanton number densities
are plotted.}
\label{fig:potfix}
\end{figure}

\begin{figure}[t]
\vspace{-0.2in}
\epsfxsize=\hsize
%\epsfbox[25 18 587 524]{pot-var.ps}
\epsffile{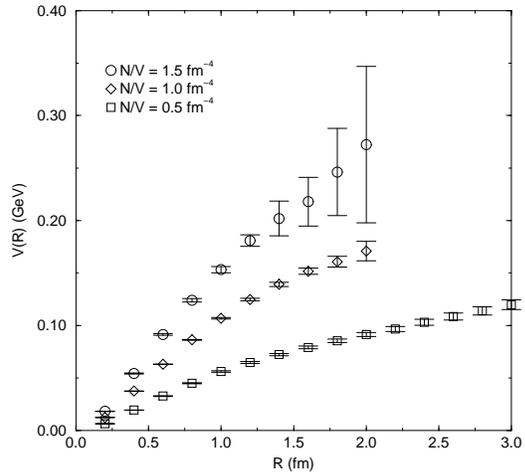}
\vspace{-0.4in}
\caption{Heavy quark potential for a variable instanton size distribution
with $\bar{\rho}=1/3fm$.  Three different instanton number densities
are plotted.}
\label{fig:potvar}
\end{figure}

Figures \ref{fig:potfix} and \ref{fig:potvar} show the heavy quark
potential out to the maximum distance we can reliably measure.  The salient
results are the following.  At short distance, the slope is identical for
both the fixed and the variable size distributions at a given $n$ and 
is proportional to $n$. 
At the value $n=1.0fm^{-4}$ of the instanton liquid model, this slope
is $\sim$ 0.1 $GeV/fm$, corresponding to 1/10 of the physical string tension.
We note that by scaling, the physical string tension would be obtained at
$n=1.0fm^{-4}$ by increasing the mean value of $\rho$ by $10^{1/4}$ to
0.59$fm$.  When $R$ is much greater than the largest instantons in the
medium, we expect $V(R)$ to approach a constant \cite{Callan78}. 
For the distribution in Eq.~(\ref{eq:size}), however, it is interesting
that the potential is essentially linear in the region of $0fm$ -- $2fm$.
Hence, even though the random instanton liquid is not strictly confining,
in the region of physical interest for hadron structure, it has a
significant linear component.

\section{ACKNOWLEDGEMENT}
The authors are grateful to Sun Microsystems for providing the 24 Gflops
E5000 SMP cluster and the Wildfire prototype system on which our
calculations were performed.

\end{document}